\documentstyle[prl,aps,psfig,twocolumn]{revtex}

\begin{document}
\twocolumn[\hsize\textwidth\columnwidth\hsize\csname
@twocolumnfalse\endcsname

\title
{
Spin and charge ordering in self-doped Mott insulators
}
\author{T. Mizokawa, D. I. Khomskii, and G. A. Sawatzky}
\address
{Solid State Physics Laboratory,
Materials Science Centre,
University of Groningen,
Nijenborgh 4, 9747 AG Groningen,
The Netherlands}
\date{\today}
\maketitle

\begin{abstract}
We have investigated possible spin and charge ordered states
in 3$d$ transition-metal oxides 
with small or negative charge-transfer energy,
which can be regarded as self-doped Mott insulators,
using Hartree-Fock calculations on $d$-$p$-type lattice models.
It was found that an antiferromagnetic state
with charge ordering in oxygen 2$p$ orbitals 
is favored for relatively large charge-transfer energy
and may be relevant for PrNiO$_3$ and NdNiO$_3$. 
On the other hand, 
an antiferromagnetic state with charge ordering 
in transition-metal 3$d$ orbitals tends to be stable
for highly negative charge-transfer energy
and can be stabilized by the breathing-type lattice 
distortion; this is probably realized in YNiO$_3$.
\end{abstract}

]

The electronic structure of 3$d$ transition metal oxides 
is described by Zaanen-Sawatzky-Allen (ZSA) scheme \cite{ZSA}
in which they are classified
into two regimes according to the relative magnitude of 
the oxygen-to-metal charge-transfer energy $\Delta$ and
the $d-d$ Coulomb interaction energy $U$. While the magnitude
of the band gap is given by $U$ in the Mott-Hubbard regime,
it is given by $\Delta$ in the charge-transfer regime 
$\Delta < U$. 
3$d$ transition-metal oxides with high valence
generally have very small or negative 
charge-transfer energy $\Delta$ and fall in a region 
which is not included in the ZSA scheme \cite{NaCuO2}.
Actually, perovskite-type 3$d$ transition-metal oxides
such as LaCu$^{3+}$O$_3$, \cite{LaCuO3} 
PrNi$^{3+}$O$_3$ \cite{PrNiO3}
and SrFe$^{4+}$O$_3$ \cite{SrFeO3} have been studied by
high energy spectroscopy and have been found
to have very small or even negative 
charge-transfer energy $\Delta$.

With small or negative $\Delta$, 
the highest part of the oxygen 2$p$ bands 
can overlap with the lowest part of the upper Hubbard
band constructed from the transition-metal 3$d$ orbitals 
so that some holes are transferred from the 3$d$ orbitals 
to the 2$p$ orbitals in the ground state. 
This state can be viewed as a self-doped state of a Mott insulator
\cite{Khomskii}
such as has recently been suggested for CrO$_2$ \cite{CrO2}. 
The properties of such system are far from clear and can be very rich.
It can be a paramagnetic metal, 
a ferromagnetic (FM) metal, and
a non-magnetic insulator similar to Kondo-insulators \cite{Khomskii}.
However, there exists another possibility which has not been 
explored until now: it may have charge ordering 
or charge density wave. It is possible that, in a self-doped
state of a Mott insulator,
holes in the oxygen 2$p$ orbitals undergo charge ordering
just like doped Mott insulators such as La$_{2-x}$Sr$_x$NiO$_4$ 
\cite{CO}.
In this letter, we study this possibility using model
Hartree-Fock (HF) calculation and
show that spin and charge ordered states
may appear in perovskites with negative $\Delta$.
Based on the calculations, we argue that this phenomena
occurs in perovskites containing Fe$^{4+}$ (CaFeO$_3$)\cite{CD} 
and Ni$^{3+}$
($R$NiO$_3$ where $R$ is a rare earth) \cite{PrNiO3mi}.
Specifically, we consider the latter system, 
properties of which, especially its strange magnetic properties
remain a puzzle until now \cite{PrNiO3,PrNiO3neutron}.

We use the multi-band $d$-$p$ model with 16 Ni sites
in which full degeneracy of the Ni 3$d$ orbitals
and the oxygen 2$p$ orbitals are taken into account 
\cite{HF}.
The Hamiltonian is given by\\
\begin{eqnarray}
H = H_p + H_d + H_{pd},
\end{eqnarray}
\begin{eqnarray}
H_p = {\displaystyle \sum_{k,l,\sigma}}
\epsilon^p_{k} p^+_{k,l\sigma}p_{k,l\sigma}
+ {\displaystyle \sum_{k,l>l',\sigma} 
V^{pp}_{k,ll'} p^+_{k,l\sigma}p_{k,l'\sigma}}
+ H.c.,
\end{eqnarray}
\begin{eqnarray}
H_d & = & \epsilon_d {\displaystyle \sum_{i,m\sigma}}
d^+_{i,m\sigma}d_{i,m\sigma}
+ u {\displaystyle \sum_{i,m}}
d^+_{i,m\uparrow}d_{i,m\uparrow}d^+_{i,m\downarrow}d_{i,m\downarrow}
\nonumber \\
& + & u' {\displaystyle \sum_{i,m \neq m'}}
d^+_{i,m\uparrow}d_{i,m\uparrow}d^+_{i,m'\downarrow}d_{i,m'\downarrow}
\nonumber \\
& + & (u'-j') {\displaystyle \sum_{i,m>m',\sigma}}
d^+_{i,m\sigma}d_{i,m\sigma}d^+_{i,m'\sigma}d_{i,m'\sigma}
\nonumber \\
& + & j' {\displaystyle \sum_{i,m \neq m'}}
d^+_{i,m\uparrow}d_{i,m'\uparrow}d^+_{i,m\downarrow}d_{i,m'\downarrow}
\nonumber \\
& + & j {\displaystyle \sum_{i,m \neq m'}}
d^+_{i,m\uparrow}d_{i,m'\uparrow}d^+_{i,m'\downarrow}d_{i,m\downarrow},
\end{eqnarray}
\begin{eqnarray}
H_{pd} = {\displaystyle \sum_{k,m,l,\sigma}} V^{pd}_{k,lm}
d^+_{k,m\sigma}p_{k,l\sigma} + H.c.
\end{eqnarray}
$d^+_{i,m\sigma}$ are creation operators for the 3$d$ electrons
at site $i$.
$d^+_{k,m\sigma}$ and $p^+_{k,l\sigma}$ are creation operators
for Bloch electrons with wave vector $k$ 
which are constructed from the $m$-th component 
of the 3$d$ orbitals and from the $l$-th component 
of the 2$p$ orbitals, respectively.
The intra-atomic Coulomb interaction between the 3$d$ electrons
is expressed using
Kanamori parameters, $u$, $u'$, $j$ and $j'$ \cite{Kanamori}.
The transfer integrals between the transition-metal
3$d$ and oxygen 2$p$ orbitals $V^{pd}_{k,lm}$ 
are given in terms of Slater-Koster parameters
$(pd\sigma)$ and $(pd\pi)$. The transfer integrals
between the oxygen 2$p$ orbitals
$V^{pp}_{k,ll'}$ are expressed by $(pp\sigma)$ and $(pp\pi)$.
Here, the ratio $(pd\sigma)$/$(pd\pi)$ is -2.16.
$(pp\sigma)$ and $(pp\pi)$ are fixed at -0.60 and 0.15,
respectively, for the undistorted lattice. 
When the lattice is distorted, the transfer integrals 
are scaled using Harrison's law. \cite{Harrison}
The charge-transfer energy $\Delta$ is defined by
$\epsilon^0_d - \epsilon_p + nU$, where
$\epsilon^0_d$ and $\epsilon_p$ are the energies of the bare
3$d$ and 2$p$ orbitals and $U$ ($=u -20/9j$) 
is the multiplet-averaged $d-d$ Coulomb interaction.
$\Delta$, $U$, and $(pd\sigma)$ for PrNiO$_3$ 
are 1.0, 7.0, and -1.8 eV, respectively, which are
taken from the photoemission study \cite{PrNiO3}. 

\begin{figure}
\psfig{figure=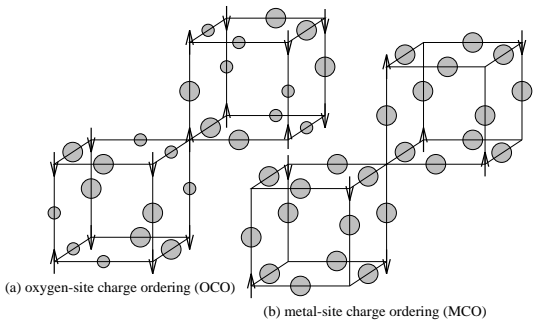,width=7cm}
\vspace{5mm}
\caption{
Schematic spin and charge orderings
(a) for the oxygen-site charge-ordered (OCO) and 
(b) for metal-site charge-ordered (MCO) states. The arrows
and the circles indicate the Ni spins 
and the oxygen 2$p$ holes, respectively.
The larger and smaller circles are for the hole-rich
and hole-poor oxygen sites, respectively.
\label{CO}
}
\end{figure}

The formally Ni$^{3+}$ (low-spin $d^7$) compounds $R$NiO$_3$
exhibit a metal-insulator transition as a function of
temperature and the size of the $R$ ion \cite{PrNiO3mi}.
Among them, PrNiO$_3$ and NdNiO$_3$ are antiferromagnetic insulators
below the metal-insulator transition temperature.
Neutron diffraction study of PrNiO$_3$ and NdNiO$_3$ 
has shown that the magnetic structure has a propagation vector
of (1/2,0,1/2) with respect to the orthorhombic unit cell or
is a up-up-down-down stacking of the ferromagnetic planes 
along the (1,1,1)-direction of the pseudocubic lattice
(see Fig. \ref{CO}(a)) \cite{PrNiO3neutron}.
In order to explain the magnetic structure, orbital 
ordering of the $x^2-y^2$ and $3z^2-r^2$ orbitals  
has been proposed because one of the $e_g$ orbitals
is occupied in the low-spin $d^7$ configuration \cite{PrNiO3neutron}.
However, previous model HF calculations have shown that 
the orbital ordered state of $x^2-y^2/3z^2-r^2$-type
has a relatively high energy,
suggesting that orbital ordering is not responsible for
the magnetic structure \cite{HF}.

The photoemission study of PrNiO$_3$ has shown that 
the charge-transfer energy $\Delta$
of PrNiO$_3$ is $\sim$ 1 eV and that the ground state
is a mixture of the $d^7$ and $d^8 \underline{L}$ configurations,
where $\underline{L}$ denotes a hole at the oxygen 2$p$ orbitals.
Since the ground state has a large amount of oxygen 2$p$ holes, 
it is also possible to describe it starting 
from the $d^8 \underline{L}$ state. In this picture, 
the system can be viewed as a self-doped Mott insulator 
and the antiferromagnetic and insulating state 
in PrNiO$_3$ and NdNiO$_3$ may be interpreted 
as a spin and charge ordered state
in the self-doped Mott insulator. Indeed, our
calculations confirmed the existence of such ordered states
which are consistent
with the neutron diffraction measurement.
They are illustrated in Fig. \ref{CO}. 
In the state shown in Fig. \ref{CO}(a), half of the oxygen sites
have more holes than the other half. 
The excess holes located at the oxygen sites
cause the ferromagnetic coupling between the neighboring two Ni spins.
Therefore, the up-up-down-down stacking of the ferromagnetic planes
along the (1,1,1)-direction is realized without orbital ordering. 
On the other hand, all the Ni sites 
have the same number of 3$d$ electrons.
Let us denote this state as an oxygen-site charge-ordered (OCO) state.
In the state shown in Fig. \ref{CO}(b), while all of the oxygen sites 
are occupied by the same amount of holes, half of the Ni sites
have more 3$d$ electrons than the other half.
This state can be called a metal-site charge-ordered (MCO) state.

\begin{figure}
\psfig{figure=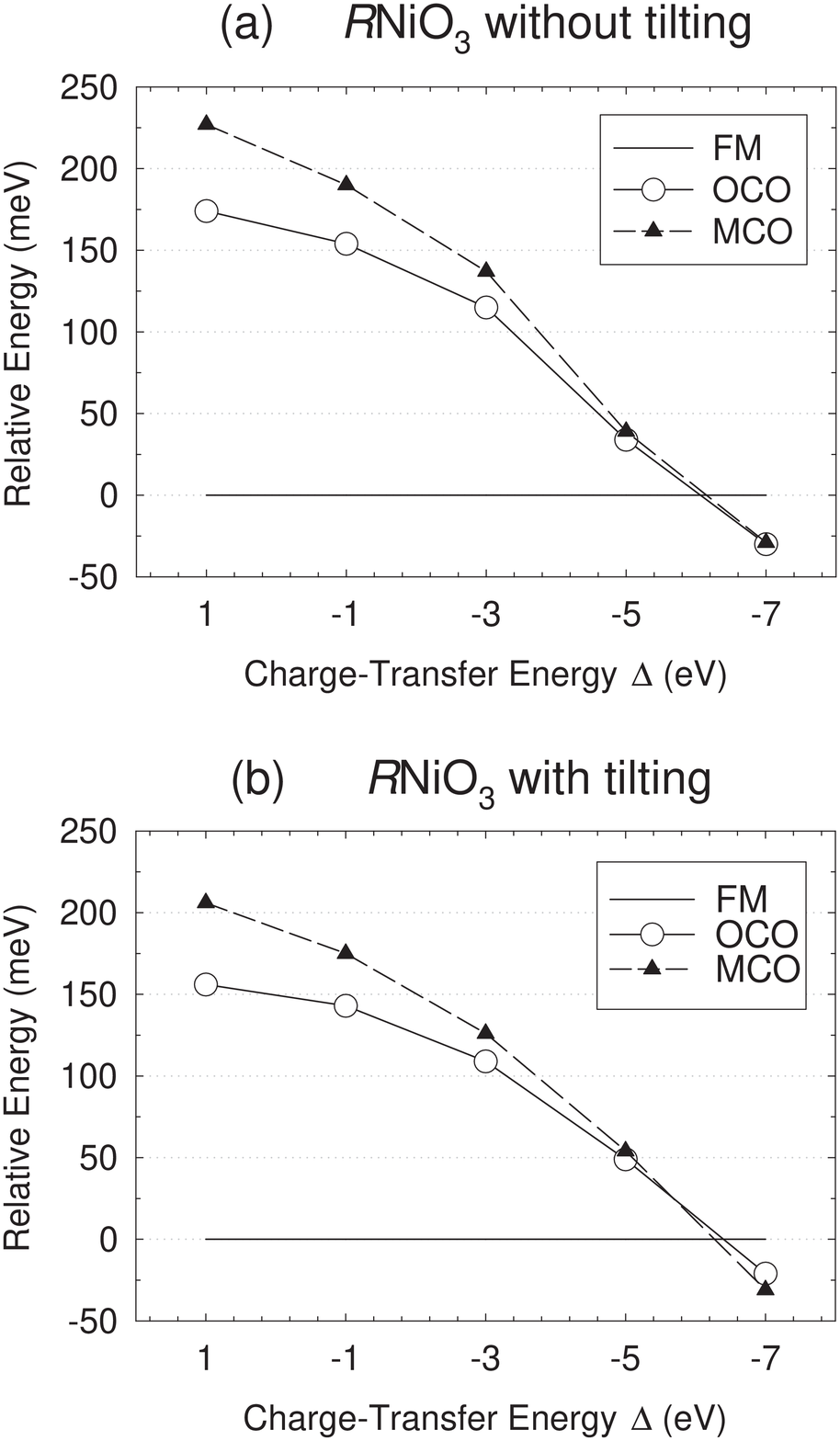,width=6cm}
\caption{
Energies of the OCO and MCO states relative
to the ferromagnetic and metallic state
as functions of the charge-transfer energy $\Delta$
(a) for the cubic perovskite lattice and (b) for the
perovskite-lattice with the orthorhombic distortion.}
\label{TE}
\end{figure}

In Fig. \ref{TE}(a), the energies of the spin and charge 
ordered states relative to the FM and metallic state
are plotted as functions of the charge-transfer energy $\Delta$
for the cubic perovskite lattice.
For $\Delta  \leq 1$ eV, the OCO and MCO states 
exist as stable solutions.  
The OCO state is lower in energy
than the MCO state for -5 eV $\leq \Delta \leq$ 1 eV.
At $\Delta$ = - 7eV, the OCO and MCO states are almost degenerate
in energy. This result indicates that, as the charge-transfer
energy $\Delta$ decreases, the MCO state becomes favored
compared to the OCO state. 
In Fig. \ref{TE}(b), the energies 
of the OCO and MCO states relative to the FM state are plotted
for the perovskite lattice with the orthorhombic distortion
which is due to the tilting of the NiO$_6$ octahedra. Here,
the tilting angle is 15$^{\circ}$ which is a typical
value found in $R$NiO$_3$ \cite{PrNiO3mi}.
At $\Delta$ = - 7eV, the MCO state
is slightly lower in energy than the OCO state, indicating
that the orthorhombic distortion or the GdFeO$_3$-type distortion
favors the MCO state. However, for $\Delta \geq$ -5 eV,
the OCO state has lower energy than the MCO state
even with the substantial distortion.
Since, in PrNiO$_3$ and NdNiO$_3$, every Ni site has the same
magnitude of the magnetic moment \cite{PrNiO3neutron}, 
it is reasonable to attribute the antiferromagnetic and 
insulating state in PrNiO$_3$ and NdNiO$_3$ to the OCO state. 
In the present model calculation without lattice distortions, 
the OCO state is higher in energy than the FM and metallic 
state for realistic $\Delta$.
However, since the charge ordering at the oxygen sites is expected to
strongly couple with a lattice relaxation, a structural modulation 
may stabilize the OCO state as discussed in the following paragraphs.

\begin{figure}
\psfig{figure=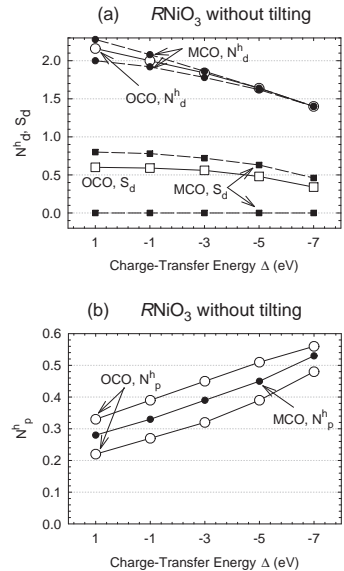,width=6cm}
\caption{
(a) Number of 3$d$ holes $N^h_d$ 
and spin $S_d$ at the Ni sites
as functions of the charge-transfer energy $\Delta$.
(b) Number of 2$p$ holes $N^h_p$ at the oxygen sites 
as functions of the charge-transfer energy $\Delta$.
}
\label{Nd}
\end{figure}

The number of 3$d$ holes $N^h_d$ and spin $S_d$ at the Ni sites
are plotted as functions of $\Delta$ in Fig. \ref{Nd}(a). In the OCO
state, $N^h_d$ is uniform at all the Ni sites. 
As $\Delta$ decreases, $N^h_d$ becomes smaller because the  
transfer of holes from the oxygen sites to the Ni sites increases.
In these solutions, $N^h_d$ are approximately two 
and the population of the $d_{x^2-y^2}$ orbital 
is the same as that of the $d_{3z^2-r^2}$ orbital,
indicating that, in the OCO state, Ni is essentially +2 
and the orbital degeneracy is lifted.
On the other hand, $N^h_d$ are 2.00 and 2.28 in the MCO state
for $\Delta$ = 1 eV. The Ni sites with $N^h_d$ of 2.00 
have the spin $S_d$ of 0.80 and are Ni$^{2+}$ as those
in the OCO state. The Ni sites with $N^h_d$ of 2.28 
have no spin and is well described by  
the $d^{8}\underline{L}^2$ state
which can hybridize with the low-spin $d^6$ state.
In this sense, the Ni sites can be viewed as 
Ni$^{4+}$-like (low-spin $d^6$) sites. 
Therefore, the MCO state is 
a kind of charge disproportionated state in which
two Ni$^{3+}$ sites are turned into the Ni$^{2+}$-like 
and Ni$^{4+}$-like sites as pointed out by Solovyev {\it et al.}
based on LDA+$U$ calculation \cite{Solvyev}.
An antiferromagnetic ordering of magnetic Ni$^{2+}$-like sites
(see Fig. \ref{CO}(b)) is also consistent with 
the neutron diffraction results \cite{PrNiO3neutron}.
As $\Delta$ decreases, the difference of $N^h_d$ between
the Ni$^{2+}$-like and Ni$^{4+}$-like sites becomes smaller
in the MCO state.
The difference almost disappears at $\Delta$ = -7 eV, where
the MCO state is almost degenerate in energy with the OCO state.
Here, it is interesting to note that the charge disproportionation
of 2Fe$^{4+}$ $\rightarrow$ Fe$^{3+}$ + Fe$^{5+}$ has been
observed in CaFeO$_3$ \cite{CD} which has highly
negative charge-transfer energy $\Delta$ \cite{SrFeO3}.
The number of 2$p$ holes $N^h_p$ at the oxygen sites
are plotted as functions of $\Delta$ in Fig. \ref{Nd}(b).
The OCO state has the hole-rich and hole-poor oxygen sites.
For $\Delta$ of 1 eV, $N^h_p$ is $\sim$ 0.33 
at the hole-rich oxygen sites 
and is $\sim$ 0.22 at the hole-poor oxygen sites.
On the other hand, in the MCO state, $N^h_p$ is
uniform at all the oxygen sites.

\begin{figure}
\psfig{figure=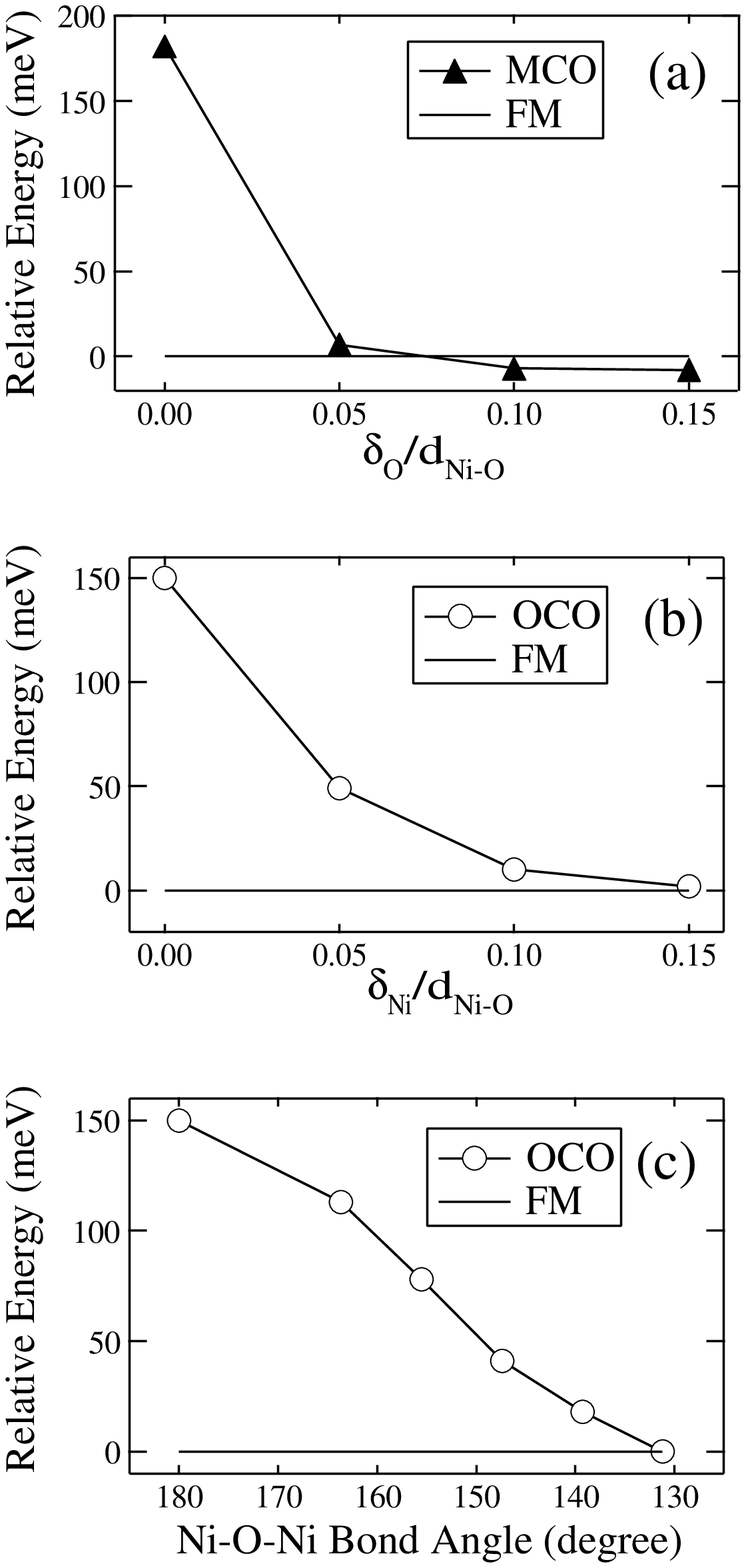,width=5cm}
\caption{
Energies of the OCO and MCO states relative
to the FM state as functions of (a) the breathing-type
distortion given by the shift of the oxygen ions, 
(b) the modulation of the bond length driven by the shift of 
the Ni ions, and (c) the modulation of the bond angle 
driven by the shift of the oxygen ions. 
$\delta_{\rm O}$ and $\delta_{\rm Ni}$ are the shifts
of the oxygen and Ni ions. $d_{\rm Ni-O}$ is the Ni-O 
bond length for the undistorted lattice.  
}
\label{T2}
\end{figure}

Recently, Medarde {\it et al.} 
have observed strong $^{16}$O-$^{18}$O 
isotope effect on the metal-insulator transition of $R$NiO$_3$,
\cite{Medarde} indicating that the electron-lattice coupling
is important in $R$NiO$_3$. 
Very recently, Alonso {\it et al.} performed
neutron diffraction studies of YNiO$_3$ and found
the breathing-type distortion which may be an indication 
of charge ordering \cite{YNiO3}. 
In Fig. \ref{T2}, the relative energies of the OCO and
MCO states compared to the ferromagnetic state are plotted 
for $\Delta$ of -1 eV as functions of the various lattice distortions
with which the charge orderings are expected to couple.
The MCO state becomes stable with the breathing-type lattice
distortion as shown in Fig. \ref{T2}(a). 
Here, $\delta_{\rm O}$ is the shift of the oxygen ions which gives
the breathing-type distortion. The MCO state becomes the lowest 
in energy for rather small distortion, indicating that the MCO state 
coupled with the breathing-type distortion is relevant for YNiO$_3$. 
The OCO state can be stabilized with the modulation of
the bond length which is a consequence of the shift of the Ni ions
as shown in Fig. \ref{T2}(b). Here, the shifts of the Ni ion are
along the (1,1,1) direction and are given by 
($\delta_{\rm Ni}$,$\delta_{\rm Ni}$,$\delta_{\rm Ni}$) and 
(-$\delta_{\rm Ni}$,-$\delta_{\rm Ni}$,-$\delta_{\rm Ni}$). 
Consequently, the Ni-O bond length for the FM coupling becomes 
shorter and that for the AFM coupling becomes longer.
Fig. \ref{T2}(c) shows that the OCO state 
is also stabilized by the modulation of 
the bond angle which is derived from the tilting 
of the octahedra and the shift of the oxygen ions. 
In this model distortion,  the Ni-O-Ni bond angle for
the FM coupling is 180 $^\circ$ and that for the AFM
coupling is smaller than 180 $^\circ$. In Fig. \ref{T2}(c),
the relative energy is plotted as a function of
the smaller Ni-O-Ni band angle. Although these distortions 
can stabilize the OCO state, we need unreasonably 
large modulations in order to make the OCO state 
lower in energy than the FM state.
We need more experimental and theoretical investigations
to identify the lattice distortion realized in 
PrNiO$_3$ and NdNiO$_3$.

In conclusion, we have studied spin and 
charge ordered states in self-doped Mott insulators
with small or negative charge-transfer energy.
It was found that two types of charge ordered states are possible:
the OCO state with charge ordering at the oxygen sites
and the MCO state with charge ordering at the transition-metal sites.
The present HF calculation without distortion has shown 
that the OCO state has lower energy than the MCO state 
for moderately small $\Delta$ and that the OCO and MCO states 
are almost degenerate for highly negative $\Delta$. 
Since, in PrNiO$_3$ and NdNiO$_3$, every Ni site has the same
magnitude of the magnetic moment \cite{PrNiO3neutron},
the antiferromagnetic and insulating state in PrNiO$_3$
and NdNiO$_3$ can be attributed to the OCO state 
of the self-doped Mott insulators. 
The OCO state in the self-doped Mott insulators is novel
in that, even without explicit doping, the spin ordering
at the transition-metal sites and the charge ordering at the
oxygen sites coexist and couple with each other just like
the spin and charge ordered states in the doped Mott insulators.
On the other hand, for YNiO$_3$, the strong breathing-type 
distortion stabilizes the MCO state. 
Here, it is interesting to note that $\Delta$ of CaFeO$_3$ 
is highly negative and can have the MCO state even 
without the strong beathing-type distortion.
The charge disproportionated state observed in CaFeO$_3$ 
may be regarded as a kind of MCO state 
in the self-doped Mott insulators. 
In $R$NiO$_3$ and CaFeO$_3$, the homogeneous state 
corresponds to an orbitally degenerate state 
($t_{2g}^6e_g^1$ for Ni$^{3+}$
and $t_{2g}^3e_g^1$ for Fe$^{4+}$). 
The charge disproportionation observed in 
CaFeO$_3$ and the OCO and MCO states for $R$NiO$_3$
may be another way to get rid of this orbital degeneracy besides
the usual cooperative Jahn-Teller (or orbital) ordering.

The authors would like to thank M. Medarde, J. Rodriguez-Carvajal, 
J. L. Garcia-Mu$\tilde{\rm n}$os, J. Matsuno, 
A. Fujimori, I. Solovyev and J. B. Goodenough
for useful discussions. This work was supported by 
the Nederlands Organization for Fundamental Research 
of Matter (FOM) and by the European Commission TRM network 
on Oxide Spin Electronics (OXSEN).

\end{document}